\pdfoutput=1

\documentclass[12pt]{article}

\usepackage{amsfonts}
\usepackage{amsmath}
\usepackage{amssymb}
\usepackage{bigints}
\usepackage{booktabs}
\usepackage[nosort]{cite}
\usepackage{color}
\usepackage{dsfont}
\usepackage{float}
\usepackage{framed}
\usepackage{graphicx}
\usepackage{indentfirst}
\usepackage{mathrsfs}
\usepackage{multirow}
\usepackage{pdflscape}
\usepackage{setspace}
\usepackage{subdepth}
\usepackage{subfig}
\usepackage{titlesec}
\usepackage[dotinlabels]{titletoc}
\usepackage{wrapfig}
\usepackage[all]{xy}
\usepackage{young}
\usepackage[vcentermath]{youngtab}
\usepackage{relsize}
\usepackage{stackengine}
\usepackage{datetime}

\usepackage{hyperref}

\numberwithin{equation}{section}

\usepackage{verbatim}

\newcommand{\be}{\begin{equation}}
\newcommand{\ee}{\end{equation}}

\newcommand{\bml}{\begin{multline}}
\newcommand{\emll}{\end{multline}}

\def\({\left(} \def\){\right)}
\def\[{\left[} \def\]{\right]}

\def\da{a^{\dagger}}

\def\al{\alpha}

\def\eps{\epsilon}

\def\v{\vec}

\def\a{\alpha}
\def\b{\bar}

\def\lam{\lambda}

\def\d{\partial}

\def\o{\omega}

\newcommand{\la}{\langle}
\newcommand{\ra}{\rangle}

\newcommand{\bea}{\begin{eqnarray}}
\newcommand{\eea}{\end{eqnarray}}

\usepackage[left=2cm,right=2cm,top=2cm,bottom=2cm]{geometry}
\titleformat{\section}{\normalfont\bfseries}{\thesection.}{4pt}{}
\titlespacing{\section}{0pt}{22pt}{6pt}

\titleformat{\subsection}{\normalfont\itshape}{\thesubsection.}{4pt}{}
\titlespacing{\subsection}{0pt}{18pt}{6pt}

\titleformat{\subsubsection}{\normalfont\itshape}{\thesubsubsection.}{4pt}{}
\titlespacing{\subsubsection}{0pt}{16pt}{6pt}

\def\ie{\begin{equation}\begin{aligned}}
\def\fe{\end{aligned}\end{equation}}

\def\bar{\overline}

\def\d{\partial}

\def\1{{\mathds 1}}

\def\o{\omega}
\def\v{\vec }

\DeclareFontShape{OT1}{cmr}{mx}{n}%
    {<->cmr10}{}
\newcommand{\mytitlefont}{\fontseries{mx}\selectfont}
\DeclareMathAlphabet{\titlemath}{OT1}{cmr}{mx}{n}

\newcommand{\bi}{\begin{itemize}}
\newcommand{\ei}{\end{itemize}}

\newcommand{\sss}{\subsubsection}

\usepackage{bm}

\begin{document}

\begin{titlepage}

\begin{center}

~\\[1cm]

{\fontsize{20pt}{0pt} \mytitlefont Wave turbulence and the kinetic equation beyond leading order  }\\[10pt]

~\\[0.2cm]

{\fontsize{14pt}{0pt}Vladimir Rosenhaus{\small $^{1}$} and Michael Smolkin{\small $^{2}$}}

~\\[0.1cm]

\it{$^1$ Initiative for the Theoretical Sciences}\\ \it{ The Graduate Center, CUNY}\\ \it{
 365 Fifth Ave, New York, NY 10016, USA}\\[.5cm]
 
 \it{$^2$ The Racah Institute of Physics}\\ \it{The Hebrew University of Jerusalem} \\ \it{
Jerusalem 91904, Israel}

~\\[0.6cm]

\end{center}

\noindent

 We derive a  scheme by which to solve the Liouville equation perturbatively in the nonlinearity, which we apply to weakly nonlinear classical field theories. 
 Our solution is  a variant of the Prigogine diagrammatic method,  and is based on an analogy  between the Liouville equation in infinite volume and  scattering in quantum mechanics, described by the Lippmann-Schwinger equation.
The motivation for our work is wave turbulence: a broad class of nonlinear classical field theories are believed to have a stationary turbulent state --- a far-from-equilibrium state, even at weak coupling. Our method provides an efficient way to derive properties of the weak wave turbulent state. A central object in these studies, which is a reduction of the Liouville equation, is the kinetic  equation, which governs  the occupation numbers of the modes. All properties of wave turbulence to date are based on the kinetic equation found at leading order in the weak nonlinearity. We explicitly obtain the  kinetic equation to next-to-leading order.

\vfill

\noindent \href{mailto:vrosenhaus@gc.cuny.edu}{vrosenhaus@gc.cuny.edu}\\ \href{mailto:michael.smolkin@mail.huji.ac.il}{michael.smolkin@mail.huji.ac.il}\\[5pt]
\end{titlepage}

\tableofcontents
~\\

\section{Introduction}

This paper will present a scheme by which to solve the Liouville equation for a classical weakly nonlinear system in infinite volume, perturbatively in the nonlinearity. Our expansion will be analogous to the construction of scattering states in quantum mechanics. We will apply our method to weakly nonlinear field theories and derive new results for the kinetic equation governing the occupation numbers of the modes, which is central to studies of weak wave turbulence \cite{Zakharov, Falkovich, Nazarenko}. Wave turbulence has been shown to occur in an incredible range of  contexts,  from surface gravity waves to waves on vibrating elastic plates, see \cite{FalkovichShavit,RS1,rodda2022experimental,FalconMordant,kochurin2022three,DURING201742, Zakharov2,   orosco2022identification,guioth2022path,galka2022emergence,Falk2021, Falkovich2022a, banks2021direct,Schlichting:2019abc, Chatrchyan:2020cxs,zhu2022direct, dematteis2022structure, david2022wave, sano2022emergent, berti2022interplay}
 for some recent work.

The Liouville equation is perhaps most familiar in the context of kinetic theory of a large number of interacting particles. Under a range of assumptions, primarily that the gas is dilute, the Liouville equation becomes the Boltzmann equation, governing the single particle phase space density. The corrections to the Boltzmann equation away from the dilute gas limit can be large and give qualitatively new effects \cite{PhysRev.139.A1763,PhysRevLett.25.1257, dorfman2015nonequilibrium}, and have been studied in a number of works \cite{Dorfman, green1956boltzmann, cohen1962generalization,zwanzig1963method, bogoliubov1960problems,brocas1967comparison, balescu1960irreversible, Prigogine, Liboff}. 

The Liouville equation can just as well be studied for interacting waves, rather than interacting particles. In the weak interaction limit it reduces to the wave kinetic equation, the wave analog of the Boltzmann equation \cite{Peierls}. In addition to the thermal state there is, surprisingly, a second stationary solution -- a far-from-equilibrium turbulent state (the Kolmogorov-Zakharov state). The properties of this turbulent state, such as the occupation numbers of the modes and the correlations between modes, are  largely unknown, except at leading order in the weak interaction. However, experience from other contexts, such as the kinetic theory of gases or  perturbative quantum field theory, suggests that going beyond leading order is possible, valuable, and can lead to qualitatively new effects.  There has been a recent push to study the wave turbulent state beyond leading order  \cite{ FalkovichShavit,RS1}, and this work is in line with this motivation. 

Long ago Prigogine observed that the Liouville equation can be viewed as a Schr\"odinger equation, and used time-dependent perturbation theory to study the approach to equilibrium, going beyond the Boltzmann equation \cite{Prigogine}. We build on Prigogine's work, with the distinction that we draw an analogy between stationary solutions of  the Liouville equation and solutions of the time-independent Schr\"odinger equation. More specifically, with a continuum of modes, the Liouville equation is analogous to  the Lippmann-Schwinger equation, used in quantum mechanics to describe scattering states. We proceed to solve the Liouville equation, formulated as a  Lippmann-Schwinger equation, perturbatively in the nonlinearity. We do this explicitly for field theories with a weak quartic interaction, to the first two orders in the nonlinearity.

The paper is organized as follows. In Sec.~\ref{sec2} we view the Liouville equation as a Schr\"odinger equation and set up the perturbation theory for the solution. In Sec.~\ref{sec3} we specialize to Hamiltonians for waves with a weak quartic interaction and work out the perturbative solution of the Liouville equation at leading order in the quartic interaction, reproducing the well-known kinetic equation for waves. In Sec.~\ref{sec4} we extend this to next-to-leading order in the interaction and,  in particular, derive the kinetic equation to next-to-leading order.  We conclude in Sec.~\ref{dis}.

\section{A perturbative solution of the  Liouville  equation} \label{sec2}
The statistical mechanics of $N$ degrees of freedom is concerned with solving for the evolution of the phase space density $\rho(J_i,\al_i,  t)$, a function of the action, $J_i$, and angle, $\al_i$, variables, with $i$ running from $1$ to $N$. The evolution is governed by the Liouville equation, 
\be \label{Liou}
\frac{\d \rho}{\d t} = \{H, \rho\}~.
\ee
One is often interested in stationary solutions, which describe the late-time behavior. For almost any interacting system, the expectation is that this stationary solution is the thermal state. Surprisingly, for many field theories there is an additional stationary solution, which is the ``wave turbulent'' state. Concretely, what is known is this: for a field theory, at weak coupling, the stationary solution of the Liouville equation is of the form, 
  \be \label{rho1}
 \rho(J) = \exp\( - \sum_p \frac{J_p}{n_p}\)~, \ \ \ \ \  \ \ \la J_p\ra \equiv \int d J\, J_p\, \rho(J) = n_p~,
 \ee
 where the sum is over all Fourier modes $p$. The thermal solution has the occupation numbers, 
 \be \label{23}
 n_p = \frac{T}{\omega_p}~,
 \ee
 which is the Rayleigh-Jeans solution, where $T$ is the temperature (this is simply the Bose-Einstein occupation number at large temperatures) and $\omega_p$ is the energy of mode $p$. For certain Hamiltonians, there is a second solution -- the wave turbulent, or Kolmogorov-Zakharov, solution -- with $\rho$ of the form  (\ref{rho1}), but with an $n_p$ that scales as a power of $p$, 
 \be
 n_p \sim p^{- \gamma}~,
 \ee
 where $\gamma$ depends on the form of the interaction. 
 This is a stationary solution which is very different from the thermal solution; it is a state that is far from equilibrium. 
 
 The thermal state at finite coupling is similar to the one at weak coupling: it is of the form (\ref{rho1}), but the $\omega_p$ appearing in $n_p$ in (\ref{23}) is replaced with the full energy of mode $p$. On the other hand, the density $\rho(J_i, \al_i)$ for the turbulent state beyond leading order in the coupling is far more interesting and elaborate: it will be a complicated function of the $J_i$ and the $\al_i$. Finding $\rho(J_i, \al_i)$ is a major problem, and this will be our goal: to systematically solve the Liouville equation, perturbatively in the coupling, for a time-independent $\rho$. 

A way of organizing the perturbation series is to notice that the Liouville  equation can be viewed as a Schr\"odinger equation. Explicitly, after multiplication by $i$, (\ref{Liou}) becomes, 
\be \label{Liou1}
i \frac{\d \rho}{\d t} = L \rho~,
\ee
where
\be \label{Liou2}
L= i \sum_j \( \frac{\d H}{\d \al_j} \frac{\d}{\d J_j} -\frac{\d H}{\d J_j} \frac{\d}{\d \al_j} \)~.
\ee
A notable difference with the Schr\"odinger equation is that  the phase space density $\rho(J_i, \al_i)$ depends on both the action and the angle variables, unlike the wavefunction which depends on either the positions or the momenta. 

We are interested in stationary solutions of the Liouville equation. This amounts to finding $\rho$ for which $L \rho =0$; or, in the language of the Schr\"odinger equation, solving the time-independent Schr\"odinger equation for eigenfunctions with zero eigenvalue. There are two kinds of eigenfunctions of the Schr\"odinger equation: bound states and scattering states. It is the scattering states that we are  interested in. In quantum mechanics one finds the scattering state eigenfunctions by solving the Lippmann-Schwinger equation. Let us recall how this works; we will then do something analogous for the Liouville equation.

\sss*{Lippmann-Schwinger equation}
The Hamiltonian in  quantum mechanics is split into a free part and an interacting part, $H = H_0 + V$, so that the Schr\"odinger equation is, 
 \be
 (H_0 + V) |\psi\ra = E |\psi\ra~.
 \ee
 The solution is formally \cite{Sakurai},
 \be \label{LS}
 |\psi\ra = |\psi_0\ra +G(E) V |\psi \ra~, \ \ \ \ G(E) = \frac{1}{E- H_0 + i\eps}~,
 \ee
 where $|\psi_0\ra$ is the solution of the free Schr\"odinger equation, $H_0 |\psi\ra = E |\psi\ra$.~\footnote{The $i\eps$ that we added sets the boundary conditions. In particular, we can view the $i\eps$ as being added to the energy and making it slightly complex, $E \rightarrow E + i \eps$. The eigenfunctions have the time evolution $e^{- i E t}$, which becomes $e^{- i E t}e^{ \eps t}$. This vanishes in the far past, which is what we want: we want the far past to have eigenfunctions of the free Schr\"odinger equation. The choice of $+ i\eps$ which we made corresponds to a retarded propagator. It means that a plane wave is ingoing and a scattered wave is outgoing. The choice of $-i\eps$ corresponds to an advanced propagator. Notice that the $i\eps$ is absent once we Fourier transform to position space in (\ref{Gpos}). If we had picked $-i\eps$ in (\ref{LS}), instead of $+ i\eps$, the exponential in (\ref{Gpos}) would have had a minus sign,  leading to a scattered wave  that is  ingoing.} Eq.~\ref{LS} is the Lippmann-Schwinger equation. Iterating the equation gives a perturbative expansion for the wavefunction, 
 \be \label{LSe}
 |\psi\ra  = |\psi_0\ra +G(E) V |\psi_0 \ra+G(E) V G(E) V |\psi_0 \ra + \ldots~.
 \ee
 It is useful to write the Lippmann-Schwinger equation in position space, 
\be
\psi(x) = \psi_0(x) + \int dx' G(x,x')V(x')\psi(x')~, \ \ \ G(x,x') = \la x| G(E)| x'\ra~, 
\ee
where the position-space wavefunction  is $ \psi(x) = \la x|\psi\ra$, and the Green's function is $G(x,x') = \la x| G(E)| x'\ra$, while the potential is taken to be local, $\la x| V| x'\ra = V(x) \delta(x-x')$. For $H_0$ corresponding to  a free particle, $H_0 = p^2/2m$, the propagator is, 
\be \label{Gpos}
G_{\pm}(x,x') = \int \frac{d^3 p}{(2\pi)^3} \frac{e^{i \v p\cdot (\v x- \v x')}}{E - \frac{p^2}{2m} \pm i\eps} = -\frac{m}{2\pi}\frac{e^{ i k |\v x - \v x'|}}{|\v x - \v x'|}~, \ \ \ \ E= \frac{k^2}{2m}~.
\ee
Finally, we note that the Lippmann-Schwinger equation applies to scattering states, which have a continuum of energy levels. This was important to us when we identified the energy of the eigenstates of both the free Hamiltonian and the interacting Hamiltonian as $E$. This is to be contrasted with bound states: the bound state energy levels are discretely spaced and change as the Hamiltonian changes.

 \sss*{Liouville equation}
 We now return to the Liouville equation (\ref{Liou1}) and split the Hamiltonian into a free part, which depends  only on the action variables, and an interacting part, 
  \be
 H = H_0(J_j) +  V(J_j, \al_j)~.
 \ee
 The Liouville operator is correspondingly $L = L_0 + \delta L$ where, 
  \bea \label{Lio1}
L_0 &=&  - i\sum_j \omega(J_j) \frac{\d }{\d \al_j}~, \ \ \ \ \ \ \ \ \  \omega(J_j) \equiv \frac{\d H_0}{\d J_j}~,\\ \label{Lio2}
 \delta L&=& i \sum_j \( \frac{\d V}{\d \al_j} \frac{\d}{\d J_j} -\frac{\d V}{\d J_j} \frac{\d}{\d \al_j} \)~.
 \eea
 
 Our goal is to solve the Liouville equation for a time-independent phase space density $\rho$. We denote such a state in ket notation as $|\Phi\ra$, so that $\rho(J_i, \al_i) = \la J_i, \al_i |\Phi\ra$. As a result of the  Liouville equation (\ref{Liou1})  $|\Phi\ra$ must satisfy, 
 \be \label{L0dL}
(L_0 + \delta L) |\Phi\ra = 0~.
\ee
In analogy with the Lippmann-Schwinger equation (\ref{LS}), the solution of (\ref{L0dL}) is,
\be \label{L2}
|\Phi\ra = |\Phi_0\ra + G \delta L |\Phi\ra~, \ \ \ \ G= \frac{1}{-L_0{+} i\eps}~,
\ee
where $|\Phi_0\ra$ solves $L_0 |\Phi_0\ra = 0$. This is just like (\ref{LS}) with $L_0$ instead of $H_0$ and with $E=0$. Here the propagator (or Green's function) $G$ is the inverse of the free Liouville operator $L_0$.  Let us look at the Green's function in more detail. Writing its matrix element as, $\la \v \al | G| \v \al'\ra = G(\v \al{-}\v \al') $, we have,
\be
L_0 G(\vec \al) =- \delta(\vec \al)~, \ \ \ \ \ \ G(\vec \al) = \sum_{\vec n} \frac{e^{i \vec n \cdot \vec \al}}{-\vec n \cdot \vec \o + i\eps}~,
\ee
where $ \delta(\vec \al)\equiv \prod_j\delta(\al_j)$. Here  $\vec n$ is a vectors of integers, and we have switched to vector notation, $\v J, \v \al$ to denote the $J_i, \al_i$. We denote the Fourier transform of the Green's function by $G(\vec n)$, 
\be \label{Gn}
G(\vec n) = \frac{1}{-\vec n \cdot \vec \o +i\eps}~.
\ee
Notice that the eigenfunctions $|\v n\ra$ of the free Liouville operator $L_0$ are plane waves, 
\be
\la \v\al| \v n\ra = \exp\( i \v n \cdot \v \a\)~, \ \ \ \ \ L_0 |\v n\ra  = \v n \cdot \v \omega  |\v n\ra~.
\ee
We could multiply this by any function $ \rho(\v J)$ of the action variables and it would continue to be an eigenfunction of $L_0$. We are particularly interested in the eigenfunctions of $L_0$ with zero eigenvalue, as they correspond to the solutions $|\Phi_0\ra$ of the free Liouville equation, 
\be \label{Phi0}
\la \v \al | \Phi_0\ra =  \exp\( i \v n \cdot \v \a\) \rho(\v J)~, \ \ \ \ \v n \cdot \v \omega =0~,
\ee
where $\rho(\v J)$ is any function of the action variables. Even though this a function of the $\v J$ variables as well, for notational simplicity we  omit writing an explicit $\v J$ on the left-hand side. 
Also, to be clear, we wrote the sum in (\ref{Lio1}) and (\ref{Lio2}) to be over a seemingly discrete index $j$, however we really have in mind that $j$ is momentum, which is continuous. For instance, the true meaning of $\v n \cdot \v \omega$ is, 
\be \label{ndoto}
\v n \cdot \v \omega \equiv \sum_in_i \o_i = \sum_p n_p \o_p= \int d^d p\,  n_p\, \o_p~.
\ee
We will continue to use the $\v n \cdot \v \omega$ notation due to its simplicity. The $\v n  \cdot \v \omega$ is evidently analogous to  energy in the discussion of the Lippmann-Schwinger equation. 

 The Liouville equation (\ref{L2}), written in the plane wave basis, is, 
\be \label{221}
\la \v n | \Phi\ra = \la \v n | \Phi_0\ra + \sum_{\v n' }G(\v n) \la \v n| \delta L |\v n'\ra \la\v n' |\Phi\ra~,
\ee
where we used that $\la \v n | G | \v n'\ra = \delta_{n, n'} G(\v n)$. The transformation between the  state in the two bases, $|\v \al\ra$ and $| \v n\ra$, is as usual through a Fourier transform, 
\be
\la \v \al | \Phi\ra = \sum_{\v n} \la \v \al | \v n\ra \la \v n| \Phi\ra  = \sum_{\v n} e^{i \v n \cdot \v \al}\la \v n | \Phi\ra~.
\ee

Just as we did for the Lippmann-Schwinger equation (\ref{LSe}), we may perturbatively expand $|\Phi\ra$ around $|\Phi_0\ra$, 
\be \label{223}
\!\!\!\!\la \v n | \Phi\ra = \la \v n | \Phi_0\ra + \sum_{\v n' }G(\v n) \la \v n| \delta L |\v n'\ra \la n' |\Phi_0\ra+
 \sum_{\v n', \v n''} G(\v n) \la \v n| \delta L |\v n'\ra G(\v n') \la \v n'| \delta L |\v n''\ra \la \v n'' |\Phi_0\ra +\ldots~.
 \ee
This is an expression that builds the solution $|\Phi\ra$ of the full Liouville equation in terms of a solution $|\Phi_0\ra$ of the free Liouville equation. Let us take a special case:  $\la \v \al |\Phi_0\ra$ which is independent of $\v \al$,~\footnote{We do this because later on we will be interested in  $\la J_r\ra = \int d J d\al\, J_r\, \la \al |\Phi_0\ra$, which will vanish unless $\la \al |\Phi_0\ra$ (\ref{Phi0}) is independent of $\al$.}
\be \label{rho0}
\la \v \al |\Phi_0\ra = \rho(\v J)~.
\ee
The Fourier transform  $\la \v n| \Phi_0\ra$ is correspondingly only nonzero for $\v n = 0$, with $\la \v n =0|\Phi_0\ra = \rho(J)$. We therefore set $\v n=0$   in (\ref{223}) so that it becomes, 
\be \label{223v2}
\la  0 | \Phi\ra = \la 0 | \Phi_0\ra +G(0) \la 0| \delta L | 0\ra \la 0|\Phi_0\ra+
 \sum_{\v n'} G(0) \la  0| \delta L |\v n'\ra G(\v n') \la \v n'| \delta L |0\ra \la 0 |\Phi_0\ra +\ldots~.
 \ee
 In the next section we will specialize these equations to a field theory with a quartic interaction. 

\section{Quartic field theory: leading order} \label{sec3}
We now turn to our Hamiltonian of interest, for a field theory with a quartic interaction. Rather than working with real fields $\phi_p$ and momentum conjugate $\pi_p$, it is common to work with complex fields $a_p$ and their complex conjugate $a_p^*$, which we will denote by $\da_p$.~\footnote{The relation between them is, as usual,  $\phi_k = \frac{1}{2\sqrt{\o_k}}(a_k + \da_k)$ and $\pi_k = \frac{i}{2\sqrt{\o_k}}(\da_k - a_k)$.} The Hamiltonian is taken to be \cite{Falkovich}, 
  \be \label{H21}
H = \sum_p \o_p \da_p a_p + \sum_{p_1, p_2, p_3, p_4}\!\! \lambda_{p_1 p_2 p_3 p_4} \da_{p_1} \da_{p_2} a_{p_3} a_{p_4}~,
\ee
where $\o_p$ is an arbitrary real function of $p$ and the coupling $\lambda_{p_1 p_2 p_3 p_4}$ is any complex function of the momenta $p_1, p_2, p_3, p_4$ which has the symmetry properties $\lam_{p_1 p_2 p_3 p_4} = \lam_{p_2 p_1 p_3 p_4} =\lam_{p_1 p_2 p_4 p_3}  = \lam_{p_3 p_4 p_1 p_2}^*$. 
 Phase space in these variables is a function of the $a_p$ and $\da_p$. It is convenient to switch to  action-angle variables, $J_p, \al_p$, 
 \be \label{32}
 a_p = \sqrt{J_p} e^{-i \al_p}~, \ \  \ \da_p = \sqrt{J_p} e^{i \al_p}~,
 \ee
so that the Hamiltonian becomes,
 \be \label{33}
 H = \sum_i \o_i J_i + \sum_{i,j,k,l} \lam_{ijkl} \sqrt{J_i J_j J_k J_l} \exp\( i (\al_i{ +} \al_j {-} \al_k{ -} \al_l)\)~,
 \ee
 where we have switched to labeling the momenta $p$ by the discrete index $i$, for notational convenience.  
The advantage of action-angle variables is that the free part of the Hamiltonian depends only on the action variables. 

Computing the Liouville operator, we have that $L_0$ is (\ref{Lio1}) with $\omega(J_i) = \omega_i$,  while $\delta L$ in (\ref{Lio2}) is,
\be
\delta L = - \sum_{i, j,k,l} \lam_{i j k l} \sqrt{J_i J_j J_k J_l}\, e^{- i \v e_{ij; k l} \cdot \v \al} \[\(\d_i {+} \d_j{ -} \d_k {-} \d_l \)+ \frac{i}{2}\(\frac{\d_{\al_i}}{J_i}+ \frac{\d_{\al_j}}{J_j}+\frac{\d_{\al_k}}{J_k}+ \frac{\d_{\al_l}}{J_l} \)\]~,
\ee
where we are using the shorthand $\d_i \equiv \d_{J_i}$. In addition, we have defined $\v e_{i j; kl} $, which is a vector that has a $-1$ in the $i$'th and $j$'th entry and a $+1$ in the $k$'th and $l$'th entry, and a zero elsewhere; in other words,
\be
-\v e_{ij; k l} \cdot \v \al = \al_i{+}\al_j {-}\al_k{-}\al_l~.
\ee 
The matrix element of $\delta L$ trivially follows, 
\be \label{Lmat}
\!\!\! \la \v n| \delta L |\v n'\ra = -\! \sum_{i, j,k,l} \lam_{i j k l} \sqrt{J_i J_j J_k J_l}\, \delta_{\v n + \v e_{ij;kl}, \v n'} \[ \(\d_i {+} \d_j{ -} \d_k {-} \d_l \)-\frac{1}{2}\(\frac{n'_i}{J_i}+ \frac{n'_j}{J_j}+\frac{n'_k}{J_k}+ \frac{n'_l}{J_l} \)\]~.
\ee
An equivalent way of writing this, which will also be useful is, 
\be \label{Lmat2}
\!\!\! \la \v n| \delta L |\v n'\ra = -\! \sum_{i, j,k,l} \lam_{i j k l}\,  \delta_{\v n + \v e_{ij;kl}, \v n'} \[ \(\d_i {+} \d_j{ -} \d_k {-} \d_l \)-\frac{1}{2}\(\frac{n_i}{J_i}+ \frac{n_j}{J_j}+\frac{n_k}{J_k}+ \frac{n_l}{J_l} \)\] \sqrt{J_i J_j J_k J_l}~,
\ee
where we used the commutation relation
$\big[\d, \sqrt{J}\, \big] =\frac{1}{2J} \sqrt{J}
$
to go between the two expressions. 

Now, using these matrix elements in (\ref{223v2}) gives, 
\be \label{37}
\!\!\!\la 0  | \Phi\ra = \la 0 |\Phi_0\ra +  G(0) \sum_{i j k l}|\lam_{ i j k l}|^2 G(\v e_{ij; k l}) \(\d_i {+} \d_j {-} \d_k {-} \d_l \)J_i J_j J_k J_l \(\d_k{+} \d_l {- }\d_i {-} \d_j\)  \la 0 | \Phi_0\ra + \ldots~,
\ee
where we used that, in order for the matrix element $\la \v 0| \delta L |\v n'\ra$ to be nonzero, the intermediate state $|\v n\ra$ has to take the form $|\b i \b j;  k  l\ra$, by which we mean the mode occupation numbers are: $n_i=n_j=-1$ and $n_k=n_l=1$, and all other $n_a=0$. 

Notice that the second term in (\ref{37}) has a $G(0)$. From the form of $G(\v n)$ given in (\ref{Gn}), we see that $G(0)$ diverges in the limit that $\eps$ goes to zero. This means that the quantity multiplying $G(0)$ must vanish, 
\be\label{38}
 \sum_{i j k l}|\lam_{ i j k l}|^2 \frac{1}{\o_i{+}\o_{j}{-}\o_k{-}\o_l {+} i\eps} \(\d_i {+} \d_j {-} \d_k {-} \d_l \)J_i J_j J_k J_l \(\d_k{+} \d_l {- }\d_i {-} \d_j\)  \rho(\v J)  = 0~,
 \ee
where we used the explicit form of $G(\v e_{i j; k l})$  and replaced $ \la 0 | \Phi_0\ra$ with $\rho(\v J)$. 
Next, we note that,
\be \label{39}
\frac{1}{x + i\eps} = P\frac{1}{x} - i \pi \delta(x)~,
\ee
where $P$ denotes principal value. 
Since the principal part of $1/(\o_k{+}\o_{l}{-}\o_i{-}\o_j )$ is odd under exchange of $i, j$ with $k,l$, whereas the other terms in (\ref{38}) are even, we are left with the delta function piece, 
\be \label{310}
 \sum_{i j k l}|\lam_{ i j k l}|^2 \delta(\o_i{+}\o_{j}{-}\o_k{-}\o_l) \(\d_i {+} \d_j {-} \d_k {-} \d_l \)J_i J_j J_k J_l \(\d_k{+} \d_l {- }\d_i {-} \d_j\)  \rho(\v J)  = 0~.
 \ee
 
 Before proceeding further with analyzing this equation, let us take stock of where we are. Our problem was to solve the Liouville equation. We wrote this in the form (\ref{L2}) resembling the Lippmann-Schwinger equation, which gives the solution as a series built off of the solution of the free Liouville equation. We took a special solution of the free Liouville equation $\la \v \al| \Phi_0\ra = \rho(\v J)$ (\ref{rho0}) which is independent of the angles $\v \al$. However we found that, rather than (\ref{L2}) giving us corrections to this, perturbatively in the coupling, as one is used to in quantum mechanics, it gave us a constraint on  $\rho(\v J)$: at leading order it must satisfy (\ref{310}). 
  
 The equation (\ref{310}) is not new: it was found by Prigogine \cite{Prigogine}. He studied what one might call the analog of time-dependent perturbation theory for the Liouville equation, whereas we are studying the analog of time-independent perturbation theory. Prigogine found,
 \be \label{311}
\frac{d\rho(\v J)}{\d t } =  -4\pi \sum_{i j k l}|\lam_{ i j k l}|^2 \delta(\o_i{+}\o_{j}{-}\o_j{-}\o_l) \(\d_i {+} \d_j {-} \d_k {-} \d_l \)J_i J_j J_k J_l \(\d_k{+} \d_l {- }\d_i {-} \d_j\)  \rho(\v J)~,
\ee
which of course agrees with (\ref{310}) if one is looking for stationary $\rho(\v J)$, as we are. Let us see how (\ref{311}) reduces to the kinetic equation. Defining the expectation values, 
\be
 \la J_i\ra = \int dJ J_i\,  \rho(J)~, \ \ \ \la J_i J_j J_k\ra = \int dJ J_i\, J_j\, J_k\, \rho(J)~,
 \ee
and multiplying (\ref{311}) by $J_r$ and integrating with respect to $\v J$ we get, 
 \be
 \frac{\d \la J_r\ra}{\d t}  =-4\pi\int d J J_r \sum_{i, j, k, l} |\lam_{i j kl}|^2 \delta(\o_i {+}\o_j {-}\o_k{-}\o_l)\(\d_i{ +}\d_j {-} \d_k{-}\d_l\) J_i J_j J_k J_l\(\d_k{+} \d_l {- }\d_i {-} \d_j\)  \rho(J)~.
 \ee
 We integrate by parts to the left (twice) and 
identify the resulting  integrals as expectation values,
  \be
\!\!\frac{\d \la J_r\ra}{\d t} = 4\pi\!\sum_{i, j, k,l}\! \( \delta_{ir} {+} \delta_{j r} {-} \delta_{kr}{-}\delta_{lr}\)\! |\lam_{i j k l }|^2 \delta(\o_i {+}\o_j {-}\o_k{-}\o_l)\!\( \la J_j J_k J_l\ra {+}\la J_iJ_k J_l\ra {-} \la J_i J_j J_l\ra{-} \la J_i J_j J_k\ra \)~,
  \ee
  where $\delta_{i j}$ is the Kronecker delta function. 
  Now, if we take $\rho(J)$ to be an exponential, 
 \be \label{314}
 \rho (J)= \frac{1}{\prod_i n_i} \exp\( - \sum_i \frac{J_i}{n_i}\)~, \ \  \ \ \ \ \la J_i\ra = n_i~, \ \ \ 
 \ee 
 then all the correlation functions factorize into one-point functions and we are left with, 
 \be
 \frac{\d n_r}{\d t} = 4\pi \sum_{i, j, k,l} \( \delta_{ir} {+} \delta_{j r} {-} \delta_{kr}{-}\delta_{lr}\) |\lam_{i j k l }|^2 \delta(\o_i {+}\o_j {-}\o_k{-}\o_l) \(\frac{1}{n_i} {+} \frac{1}{n_j} {-}\frac{1}{n_k}{-}\frac{1}{n_l}\) n_i n_j n_k n_l~.
 \ee
 This is the standard (leading order) kinetic equation for waves \cite{Falkovich}.

  Using our perturbative (in the coupling) solution of the Liouville equation analog of the Lippmann-Schwinger equation, we can go beyond the leading order kinetic equation, and compute the kinetic equation to any order in the coupling. Generalizing what we just did, from (\ref{221}) we have that, 
  \be 
\la \v 0 | \Phi\ra = \la \v 0| \Phi_0\ra +G(\v 0) \sum_{\v n' } \la \v 0| \delta L |\v n'\ra \la n' |\Phi\ra~.
\ee
Since $G(\v 0)$ is divergent, it must be the case that, 
\be
 \sum_{\v n' } \la \v 0| \delta L |\v n'\ra \la n' |\Phi\ra = 0~.
 \ee
 We will refer to this as the master equation. 
 We may multiply this by $J_r$ and integrate over $\v J$, and  identify the resulting quantity as proportional to $ \frac{\d n_r}{\d t}$, 
  \be \label{318}
 \frac{\d n_r}{\d t} \propto \int dJ\, J_r\, \sum_{\v n' } \la \v 0| \delta L |\v n'\ra \la n' |\Phi\ra~.
 \ee
This is the kinetic equation: for a stationary state, both sides vanish, as should be the case. 
We therefore have a clear procedure for computing the kinetic equation to any order in the interaction strength: we perturbatively solve (\ref{L2}) for $|\Phi\ra$ with the initial condition that $|\Phi_0\ra$ is angle independent  i.e, $\la \v n |\Phi_0\ra$ is nonzero for $\v n =0$, and insert the result into  (\ref{318}). This is what we will do in the next section. \\

We end this section with a comment. Perturbatively solving the Liouville equation is equivalent to perturbatively computing integrals of motion, see \cite{Polyakov, Gurarie, Gurarie95}. The connection with the Lippmann-Schwinger equation gives a systematic and unambiguous scheme by which to compute the kinetic equation to any order in the nonlinearity. 
It is important that we are studying field theories in the continuum, which have an infinite number of degrees of freedom. One might have worried that once one goes away from the free limit, integrability will be broken (perturbation theory will break down) and there will be no conserved quantities. However, since we have a field theory, any function $\rho(\v J)$ is conserved for the free theory; once interactions are turned on most functions indeed stop being conserved, but some remain conserved: those that satisfy the equations, such as (\ref{38}), that our perturbation theory forces on us.

\section{Quartic field theory: next-to-leading order} \label{sec4}
In this section we continue the discussion in the previous section, but go to next-to-leading order in the nonlinearity. To recap: our goal is to solve the Liouville equation for the phase space distribution $\Phi(\v J, \v \al)$, which is a function of the action $J_i$ and angle  $\al_i$ variables of all the modes, which have been grouped into vectors, $\v J$ and $\v \al$. We work perturbatively in the interaction, starting with the solution $\Phi_0(\v J, \v \al)$ of the Liouville equation for a free Hamiltonian. 

We are particularly interested in the case in which $\Phi_0$ is independent of the angle. It is convenient to change basis, from angles $\v \al$ to the Fourier transform variables of occupation numbers $\v n$. The $\Phi_0$ that is $\v \al$ independent then only depends on the zero mode, $\v n = 0$. Writing $\Phi_0$ in ket notation, it is  $\la \v n=0|\Phi_0\ra$. Going back to our perturbative expansion (\ref{223v2}) for $\la  0 | \Phi\ra $, we add the next-to-leading order term,
\be
\la  0 | \Phi\ra = \la 0 | \Phi_0\ra +G(0)\big( \(\delta \mathcal{L}\)_{\text{first}} + \(\delta \mathcal{L}\)_{\text{second}} + \ldots \big)~,
\ee
where we have dropped the $\la 0|\delta L |0\ra$ term, since it vanishes for our $\delta L$,  and we defined,
\bea \label{dLf}
\(\delta \mathcal{L}\)_{\text{first}} &=& \sum_{\v n'}\la \v 0| \delta L |\v n'\ra G(\v n') \la \v n'| \delta L |0\ra \la 0 |\Phi_0\ra \\ \label{dLs}
 \(\delta \mathcal{L}\)_{\text{second}} &=&\sum_{\v n', \v n''}  \la \v 0| \delta L |\v n'\ra G(\v n') \la \v n'| \delta L|\v n''\ra G(\v n'') \la \v n'' |\delta L   |0\ra \la 0 |\Phi_0\ra~.
\eea
As argued in the previous section, since $G(0)$ diverges we must have the master equation,
\be \label{ME}
 \(\delta \mathcal{L}\)_{\text{first}} + \(\delta \mathcal{L}\)_{\text{second}} = 0~,
\ee
to third order in the coupling. 

In the previous section we computed the first term $\(\delta \mathcal{L}\)_{\text{first}} $, see (\ref{310}). Here we will compute the next order term, $\(\delta \mathcal{L}\)_{\text{second}} $. Before proceeding, it is useful to streamline our notation and give graphical representations of the terms that appear. We will label modes by $1,2,3,\ldots$ instead of $i, j, k,\ldots$. 
In addition, we can make use of the specific form of $\delta L$ for a field theory with a quartic interaction in order to specify which states can  appear in the perturbative expansion. Namely, $\delta L$ acts by ``creating'' two particles and ``destroying'' two other particles, and its matrix element was  given earlier in (\ref{Lmat}).
We have  that $\(\delta \mathcal{L}\)_{\text{first}} $ (\ref{dLf}) reduces to, 
\be \label{42}
\(\delta \mathcal{L}\)_{\text{first}} = \sum_{1,\ldots, 4} G(\v e_{12; 34})\, \la \v 0| \delta L| \b 1 \b 2;  3 4\ra\la \b 1 \b 2; 3 4|\delta L |0\ra  \la 0 |\Phi_0\ra~.
\ee
Our notation is such that when we write $| \b 1 \b 2;  3 4\ra$ we mean a state in which modes $1$ and $2$ have occupation number $-1$ and modes $3$ and $4$ have occupation number $1$. 
\begin{figure}[t]
\centering
\includegraphics[width=1.5in]{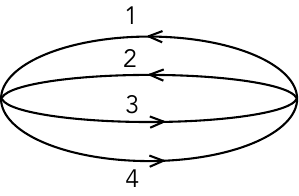}
\caption{A diagrammatic representation of $\(\delta \mathcal{L}\)_{\text{first}} $ in (\ref{42}). We start with vacuum $|0\ra$. The interaction $\delta L$, which occurs at the left vertex, turns this into the state $|\b 1 \b 2; 3 4\ra$. The two ``particles'' and two ``holes''  each propagate to the right. The interaction $\delta L$ then transforms the state back into the vacuum $|0\ra$.    } \label{Ptree}
\end{figure}
We may represent (\ref{42}) graphically, as shown in Fig.~\ref{Ptree}. One can imagine a vertical dashed line running through the center of the figure which represents the intermediate state  $| \b 1 \b 2;  3 4\ra$. The arrows on the lines indicate the direction of flow of occupation number: an arrow pointing to the right corresponds to occupation number plus one, while  an arrow pointing to the left corresponds to occupation number minus one. In this new notation (\ref{42}) in explicit form is 
\be  \label{46}
\(\delta \mathcal{L}\)_{\text{first}} = \sum_{1,\ldots, 4} G(\v e_{12; 34}) \lam_{1234}\lam_{3412} \(\d_1{+}\d_2{-}\d_3{-}\d_4\) J_1 J_2  J_3 J_4\\
\(\d_3{+}\d_4{-}\d_1{-}\d_2\) \la 0 |\Phi_0\ra~.
\ee

Let us now look at the next order term $ \(\delta \mathcal{L}\)_{\text{second}} $ (\ref{dLs}). To compute this we need to enumerate all the possible intermediate states $| \v n'\ra$ and $|\v n''\ra$. The most efficient way to do this is with the aid of diagrams describing the process. The diagrams will  have three vertices, one for each $\delta L$. There are four diagrams to consider. Two of the diagrams are shown below, in Fig.~\ref{Ploop}. 
\begin{figure}[h]
\centering
\subfloat[]{\includegraphics[width=1.5in]{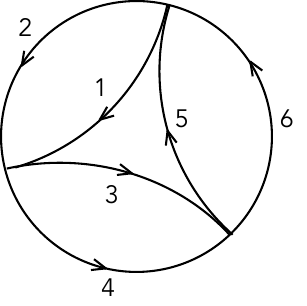}} \ \ \  \ \  \ \  \ \ \ \  \ \ \ 
\subfloat[]{\includegraphics[width=1.5in]{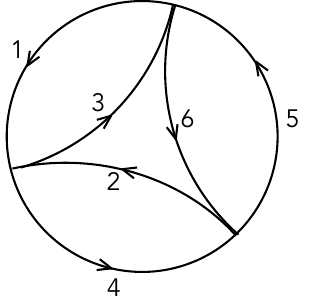}}
\caption{A diagrammatic representation of the contributions to $ \(\delta \mathcal{L}\)_{\text{second}} $. } \label{Ploop}
\end{figure} 
We also need to add the same diagrams, with all the arrows reversed. These are shown below,  in Fig.~\ref{PloopR}. 
\begin{figure}[h]
\centering
\subfloat[]{\includegraphics[width=1.5in]{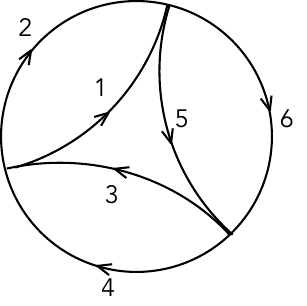}} \ \ \  \ \  \ \  \ \ \ \  \ \ \ 
\subfloat[]{\includegraphics[width=1.5in]{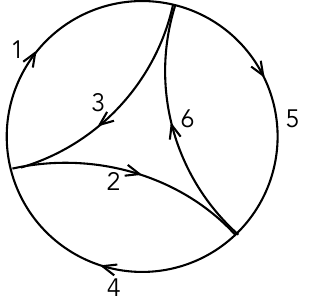}}
\caption{The same diagrams as in Fig.~\ref{Ploop}, but with all arrows reversed,  also contribute  to $ \(\delta \mathcal{L}\)_{\text{second}} $.} \label{PloopR}
\end{figure} 
We split $ \(\delta \mathcal{L}\)_{\text{second}} $ into 
\be \label{47}
 \(\delta \mathcal{L}\)_{\text{second}}  =  \(\delta \mathcal{L}\)_{\text{second}}^{a}+ \(\delta \mathcal{L}\)_{\text{second}}^{b}~,
 \ee
 where $ \(\delta \mathcal{L}\)_{\text{second}}^{a}$ denotes the contribution that is the sum of the diagrams Fig.~\ref{Ploop}(a) and Fig.~\ref{PloopR}(a), and  $ \(\delta \mathcal{L}\)_{\text{second}}^{b}$ denotes the contribution that is the sum of the diagrams Fig.~\ref{Ploop}(b) and Fig.~\ref{PloopR}(b).
Let us start
with Fig.~\ref{Ploop}(a). This diagram  can be cut in two different ways, producing two distinct intermediate states, see Fig~\ref{Ploopa}.
\begin{figure}[h]
\centering
\renewcommand{\thesubfigure}{\roman{subfigure}}
\subfloat[]{\includegraphics[width=1.3in]{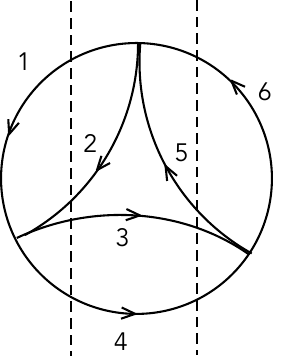}} \ \ \  \ \  \ \  \ \ \ \  \ \ \ 
\subfloat[]{\includegraphics[width=1.3in]{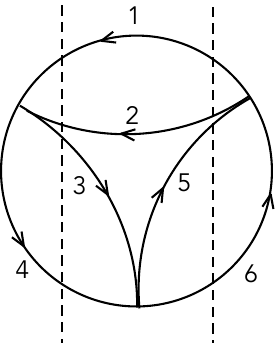}}
\caption{The diagram shown in Fig.~\ref{Ploop}(a). We use a vertical dashed line to represent an insertion of an intermediate state. Depending on the two ways in which we rotate Fig.~\ref{Ploop}(a), we produce different intermediate states, as shown here in (i) and (ii). In (i) the first (leftmost) intermediate state that the vertical dashed line crosses is $ |\b 1 \b 2; 34\ra $ and the second is $ |\b 5 \b 6; 34\ra$. In (ii) the first state is also $ |\b 1 \b 2; 34\ra $, but the second is $ |\b 1 \b 2; 56\ra$.} \label{Ploopa}
\end{figure} 

We start with  Fig.~\ref{Ploopa}(i), which corresponds to a sequence of states $|0\ra \rightarrow |\b 1 \b 2; 34\ra \rightarrow |\b 5 \b 6; 34\ra \rightarrow |0\ra$. Its contribution is, 
\be \label{43}
\sum_{1,\ldots, 6} G(0) G(\v e_{12; 34}) G(\v e_{56; 34})\, \la  0| \delta L| \b 1 \b 2;  3 4\ra \la \b 1 \b 2; 3 4|\delta L |\b 5 \b 6; 34\ra  \la \b 5 \b 6; 34| \delta L |0\ra  \la 0 |\Phi_0\ra~.
\ee
Let us evaluate this explicitly, making use of the matrix elements (\ref{Lmat}). We get, 
\bml\label{44}
-\sum_{1,\ldots, 6} G(0) G(\v e_{12; 34}) G(\v e_{56; 34}) \lam_{1234}\lam_{5612}\lam_{3456} \(\d_1{+}\d_2{-}\d_3{-}\d_4\) J_1 J_2  \sqrt{J_3 J_4 J_5 J_6}\\
\[ \(\d_5{+}\d_6{-}\d_1{-}\d_2\) +\frac{1}{2}\(\frac{1}{J_5}{+}\frac{1}{J_6}\)\]
 \sqrt{J_5 J_6 J_3 J_4}\(\d_3{+}\d_4{-}\d_5{-}\d_6\) \la 0 |\Phi_0\ra~,
\end{multline}
where we used the matrix element in the form of (\ref{Lmat2}) for  $\la \v 0| \delta L| \b 1 \b 2;  3 4\ra$, and in the form of (\ref{Lmat}) for   $ \la \b 1 \b 2; 3 4|\delta L |\b 5 \b 6;34\ra$ and $ \la \b 5 \b 6;34| \delta L |0\ra $. 
It is convenient to commute through the $J_5$ and $J_6$ so as to combine them. Using $\[\d, \sqrt{J}\] =\frac{1}{2J} \sqrt{J}
$ this gives,
\bml \label{45}
-\sum_{1,\ldots, 6} G(0) G(\v e_{12; 34}) G(\v e_{56; 34}) \lam_{1234}\lam_{5612}\lam_{3456} \(\d_1{+}\d_2{-}\d_3{-}\d_4\) J_1 J_2  J_3 J_4\\
\(\d_5{+}\d_6{-}\d_1{-}\d_2\)J_5J_6
\(\d_3{+}\d_4{-}\d_5{-}\d_6\) \la 0 |\Phi_0\ra~.
\end{multline}
Next, we note that from   Fig.~\ref{PloopR}(a) we will have an identical contribution, but with all the arrows reversed. Rather than redo the computation, it is useful to understand how to get the result by a symmetry transformation. 

\sss*{Particle-hole symmetry}

Let us understand what it means to have the arrows run in the opposite direction. For the couplings, it means $\lam_{i j k l}\rightarrow \lam_{k l i j} = \lam^*_{i j kl}$. It also means that the sign of the angles is reversed, $\v \al \rightarrow - \v \al$. In terms of $G(\v \al)$, reversing the sign of $\v \al$ corresponds in Fourier space to sending $G(\v n)$ to $G(- \v n)$. Since with our notation our occupation numbers are either plus or minus one, reversing arrows means  $G(\v e_{i j; k l}) \rightarrow G(\v e_{k l; i j}) = - G^*(\v e_{i j;kl})$.

 Now, to see the effect of reversing the arrows on the matrix elements: we are swapping particles and holes, i.e $|\b i \b j; k l\ra \rightarrow |\b k \b l; i j\ra$. We also need to understand the effect of swapping the arrows on the explicit form of the matrix elements.  In terms of the variables, we see from (\ref{Lmat}) that flipping the sign of $\v \a$ corresponds to $\v J\rightarrow - \v J$.

There is an easier way of understanding the mapping: it follows from a symmetry of the Hamiltonian (\ref{33}). In particular, the Hamiltonian is invariant under~\footnote{We have introduced the notation that if one of the indices of the coupling appears with a minus sign in the first two slots, it means the same as the index appearing without a minus sign in one of the last two slots, and likewise if an index appears with a minus sign in the last two slots, it means the same as the index appearing without a minus sign in one of the first two slots, i.e.  $ \lam_{1 -2 -3 4} \equiv \lam_{1324}$ and $ \lam_{-1 -2 -3 -4}  = \lam_{34 12}$. }
\be \label{Hsym1}
\o_i \rightarrow -\o_i~, \ \ \ J_i \rightarrow - J_i~, \ \ \ \al_i \rightarrow  - \al_i~, \ \ \ \lam_{1234} \rightarrow \lam_{-1 -2 -3 -4} \equiv \lam_{1234}^*~, \ \ \ \ i =1,2,3,4~.
\ee
It is also invariant under a flip in sign of just two of the indices, 
\be
\o_i \rightarrow -\o_i~, \ \ \ J_i \rightarrow - J_i~, \ \ \ \al_i \rightarrow  - \al_i~, \ \ \ \lam_{1234} \rightarrow  \lam_{1 -2 -3 4} \equiv \lam_{1324}~, \ \ \ \ i =2,3~.
\ee
\vspace{-.2cm}

Now, returning to (\ref{45}) and performing the described transformation gives the complex conjugate of (\ref{45}), with a minus sign. Summing the two gives what (with foresight) we denote as $\frac{1}{2}  \(\delta \mathcal{L}\)_{\text{second}}^{a}$ where,
\bml  \label{12mLa}
  \(\delta \mathcal{L}\)_{\text{second}}^{a} = -4i\sum_{1,\ldots, 6} \text{Im}\( G(\v e_{12; 34}) G(\v e_{56; 34}) \lam_{1234}\lam_{5612}\lam_{3456}\) \(\d_1{+}\d_2{-}\d_3{-}\d_4\) J_1 J_2  J_3 J_4\\
\(\d_5{+}\d_6{-}\d_1{-}\d_2\)J_5J_6
\(\d_3{+}\d_4{-}\d_5{-}\d_6\) \la 0 |\Phi_0\ra~.
\end{multline}
We also need to include the diagram in Fig.~\ref{Ploopa}(ii). We notice that it is the same as Fig.~\ref{Ploopa}(i) with $1,2 \leftrightarrow 3,4$ and the arrows going in the opposite direction. But  $1,2,3,4$ are just dummy variables, so for that contribution we can simply change variables $1,2\leftrightarrow 3,4$. So we just need to have the arrows go in the opposite direction. But to get (\ref{12mLa}) we already took the arrows to go in both directions, so this gives back the same thing:  $\frac{1}{2}  \(\delta \mathcal{L}\)_{\text{second}}^{a}$. In total we thus have  $ \(\delta \mathcal{L}\)_{\text{second}}^{a}$.

Next we turn to the diagram in Fig.~\ref{Ploop}(b). We notice  Fig.~\ref{Ploop}(b)  can be obtained from Fig.~\ref{Ploop}(a) by swapping $2$ and $3$ and then reversing the arrows on $2,3,6$. We must also multiply by a combinatorial factor of $4$ (because $1$ and $2$ can be exchanged, and $3$ and $4$ can be exchanged). Performing this transformation, $2\rightarrow - 3$, $3 \rightarrow - 2$, and $6\rightarrow - 6$, on  $ \(\delta \mathcal{L}\)_{\text{second}}^{a} $ we get 
$\(\delta \mathcal{L}\)_{\text{second}}^{b}$, 
\bml  \label{12mLb}
  \(\delta \mathcal{L}\)_{\text{second}}^{b} = 16i\sum_{1,\ldots, 6} \text{Im}\( G(\v e_{12; 34}) G(\v e_{52;46}) \lam_{1234}\lam_{5316}\lam_{4625}\) \(\d_1{+}\d_2{-}\d_3{-}\d_4\) J_1 J_2  J_3 J_4\\
\(\d_5{+}\d_3{-}\d_1{-}\d_6\)J_5J_6
\(\d_4{+}\d_6{-}\d_2{-}\d_5\) \la 0 |\Phi_0\ra~.
\end{multline}
We have now computed all the terms appearing in the master equation (\ref{ME}) to third order in the coupling. 

\sss*{Kinetic equation}
As we established in the previous section, the kinetic equation is found by integrating the master equation,
\be \label{415}
\frac{\d n_r}{\d t} = -4 i \int d J\, J_r\, \( \(\delta \mathcal{L}\)_{\text{first}} + \(\delta \mathcal{L}\)_{\text{second}} \)~.
\ee
This is straightforward to carry out; the details are in Appendix~\ref{MtoK}. The result for the first order term is, 
\be
\int d J\, J_r\,  \(\delta \mathcal{L}\)_{\text{first}}  = -\sum_{1,\ldots, 4}G(\v e_{12; 34}) \lam_{1234}\lam_{3412}  (\delta_{1r}{+}\delta_{2r} {-}\delta_{3r} {-} \delta_{4r}) \prod_{i=1}^4 n_i\,
\Big(\frac{1}{n_1}{+}\frac{1}{n_2} {-}\frac{1}{n_3} {-}\frac{1}{n_4}\Big)~,
\ee
and for the second order term is, 
\bml
\int d J\, J_r\,  \(\delta \mathcal{L}\)_{\text{second}}^a= -4i\sum_{1,\ldots, 6} \! \text{Im}\( G(\v e_{12; 34}) G(\v e_{56; 34}) \lam_{1234}\lam_{5612}\lam_{3456}\) (\delta_{1r}{+}\delta_{2r} {-}\delta_{3r} {-} \delta_{4r})\\
\prod_{i=1}^6 n_i
\Big(\frac{1}{n_1}{+}\frac{1}{n_2} \Big) 
\Big(\frac{1}{n_3}{+}\frac{1}{n_4} {-}\frac{1}{n_5} {-}\frac{1}{n_6}\Big)~,
\end{multline}
and  something similar for $\int d J\, J_r\,  \(\delta \mathcal{L}\)_{\text{second}}^b$ which is  obtained by an index interchange. Now, for $G(\v e_{12;34})$ recall its definition (\ref{Gn}), 
\be \label{Gsplit}
G(\v e_{12;34}) = \frac{1}{\o_1{+}\o_2{-}\o_3{-}\o_4 {+} i\eps} = P \frac{1}{\o_1{+}\o_2{-}\o_3{-}\o_4 } - i \pi \delta(\o_3{+}\o_4{-}\o_1{-}\o_2)~,
\ee
where we split  the term into a principal part and a delta function, using (\ref{39}). A potentially more useful form of the kinetic equation is one in which one makes use of (\ref{Gsplit}) to write the terms, grouping them by the number of delta functions. Doing this, along with some permutations of the indices to combine and simplify terms (see Appendix~\ref{MtoK} for details) we get for the kinetic equation, 
\bml \label{KEQ}
\frac{d n_1}{dt} = 16\pi \sum_{2,3, 4} \delta(\o_{12;34}) \lam_{1234}^2  \prod_{i=1}^4 n_i\, \Big( \frac{1}{n_1} {+} \frac{1}{n_2}{-}\frac{1}{n_3} {-} \frac{1}{n_4} \Big)\\
\[ 1 + 4 \sum_{5,6} \frac{\lam_{1256}\lam_{5634}}{\lam_{1234}} \frac{n_5 {+} n_6}{\o_{12;56}}+ 16\sum_{5,6} \frac{\lam_{1635}\lam_{2546}}{\lam_{1234}} \frac{n_6 {-} n_5}{\o_{46;25}}\]~,
\end{multline}
where  here we have taken the couplings to be real and we defined $\o_{ij; kl} \equiv \o_i {+} \o_j {-} \o_k {-} \o_l$, and when we write $1/\o$ we really mean the principal part of $1/\o$. This matches the next-to-leading order kinetic equation we found  in \cite{RS1}.

\section{Discussion} \label{dis}

We have studied classical field theories with weak nonlinearity. A fundamental object is the kinetic equation, describing the time evolution of the occupation number $n_k$ of a mode of momentum $k$. We presented a scheme to find the kinetic equation perturbatively in the nonlinearity, and found it explicitly to next-to-leading order for a field theory with a quartic interaction. Our method amounts to perturbatively solving the Liouville equation, which we did by analogy with the Lippmann-Schwinger equation used in scattering in quantum mechanics. 

The expansion we found is analogous to perturbation theory in many-body quantum mechanics or quantum field theory. In particular, the phase space density is a function of the action and angle variables of every mode $p$. The effect of a quartic interaction on the Liouville operator is that it adds one unit of angular dependence for some modes $p_1$ and $p_2$, and subtracts one unit for modes $p_3$ and $p_4$; it is as if the interaction creates particles of momenta $p_1$ and $p_2$ and destroys particles of momenta $p_3$ and $p_4$. In this way, each term in the perturbative expansion is represented by a Feynman diagram. Our seed for  the perturbative expansion was a phase space density that only depends on the action variables. Order by order in perturbation theory, angular dependence for an increasing number of modes is acquired. 

More broadly, the standard approach to solving the Liouville equation is through  time-dependent perturbation theory \cite{Prigogine}. In contrast, our method is like time-independent perturbation theory and works by analogy with scattering, which is far simpler. 
This approach may be useful in other contexts. 

Rather than doing a phase space average, an alternative way to derive the kinetic equation is to  introduce external Gaussian-random forcing and dissipation, perform an average over the forcing, and at the end take both the forcing and dissipation to zero while maintaining a fixed ratio. This is what we did in earlier work \cite{RS1}. The method in this paper is technically far simpler, although the method of \cite{RS1} is perhaps conceptually simpler. 

One motivation for studying weakly nonlinear field theories is the existence of turbulent cascades (the Kolmogorov-Zakharov state). The state is found by looking for stationary solutions of the wave kinetic equation. This is always done for the part of the kinetic equation that is leading order in the nonlinearity. It will be interesting to see how accounting for the next-to-leading order term in the kinetic equation modifies the turbulent state.   

Although we focused on the occupation numbers of modes, the method presented here allows one to compute much more: it systematically gives the entire state, as a function of all the action and angle variables, perturbatively in the coupling of the nonlinear term. An important problem is to decide which quantity in particular provides an insightful characterization of the turbulent state, and  to then compute it.

\sss*{Acknowledgments} 
We are indebted to G.~Falkovich for  making us aware of the Prigogine diagrammatic method  and for  valuable discussions. We thank M.~Srednicki for helpful discussions. The work of VR is supported in part by NSF grant 2209116 and by the ITS through a Simons grant. 
The work of MS is supported by the  Israeli Science Foundation Center of Excellence (grant No.~2289/18). This work was also supported in part by BSF grant 2022113.

\appendix

\section{From the master equation to the kinetic equation} \label{MtoK}
In this appendix we fill in some of details involved in getting the kinetic equation to next-to-leading order. In particular, we start with the kinetic equation (\ref{415}), make use of $\(\delta \mathcal{L}\)_{\text{first}}$ in (\ref{46}) and $ \(\delta \mathcal{L}\)_{\text{second}}$ in (\ref{47}), (\ref{12mLa}), and (\ref{12mLb})  to obtain the final form of the kinetic equation in (\ref{KEQ}). 

We first look  at the integral of $ \(\delta \mathcal{L}\)_{\text{first}}$, given   in (\ref{46}), 
\be
\int d J\, J_r\,  \(\delta \mathcal{L}\)_{\text{first}}  = \int d J\, J_r  \sum_{1,\ldots, 4}\! G(\v e_{12; 34}) \lam_{1234}\lam_{3412} \(\d_1{+}\d_2{-}\d_3{-}\d_4\) J_1 J_2  J_3 J_4\\
\(\d_3{+}\d_4{-}\d_1{-}\d_2\)  \rho(\v J)
\ee
where $\la 0|\Phi_0\ra \equiv \rho(\v J)$. As shown in Sec.~\ref{sec3},  we repeatedly integrate by parts to the left and then use  the exponential distribution for $\rho(\v J)$, as given in (\ref{314}), so that this turns into, 
\be
\int d J\, J_r\,  \(\delta \mathcal{L}\)_{\text{first}}  = -\sum_{1,\ldots, 4}G(\v e_{12; 34}) \lam_{1234}\lam_{3412}  (\delta_{1r}{+}\delta_{2r} {-}\delta_{3r} {-} \delta_{4r}) \prod_{i=1}^4 n_i\,
\Big(\frac{1}{n_1}{+}\frac{1}{n_2} {-}\frac{1}{n_3} {-}\frac{1}{n_4}\Big)~.
\ee
Next  we look at the integral of $ \(\delta \mathcal{L}\)_{\text{second}}^a$, given  in (\ref{12mLa}),
\bml 
\int d J\, J_r\,  \(\delta \mathcal{L}\)_{\text{second}}^a= -4i\int d \v J\, J_r\, \sum_{1,\ldots, 6} \text{Im}\( G(\v e_{12; 34}) G(\v e_{56; 34}) \lam_{1234}\lam_{5612}\lam_{3456}\)  \(\d_1{+}\d_2{-}\d_3{-}\d_4\)\\ J_1 J_2  J_3 J_4
 \(\d_5{+}\d_6{-}\d_1{-}\d_2\) J_5J_6
\(\d_3{+}\d_4{-}\d_5{-}\d_6\) \rho(\v J)~.
\end{multline}
We repeatedly integrate by parts to the left so that this turns into, 
\bml
\int d J\, J_r\,  \(\delta \mathcal{L}\)_{\text{second}}^a= -4i\int d \v J\sum_{1,\ldots, 6} \text{Im}\( G(\v e_{12; 34}) G(\v e_{56; 34}) \lam_{1234}\lam_{5612}\lam_{3456}\) (\delta_{1r}{+}\delta_{2r} {-}\delta_{3r} {-} \delta_{4r})\\ J_1 J_2  J_3 J_4 J_5J_6
\Big(\frac{1}{J_1}{+}\frac{1}{J_2} \Big) 
\Big(\frac{1}{J_3}{+}\frac{1}{J_4} {-}\frac{1}{J_5} {-}\frac{1}{J_6}\Big)  \rho(\v J)~.
\end{multline}
Evaluating using an exponential distribution for $\rho(\v J)$, as given in (\ref{314}), turns all the $J_i$ into $n_i$, 
\bml \label{49}
\int d J\, J_r\,  \(\delta \mathcal{L}\)_{\text{second}}^a= -4i\sum_{1,\ldots, 6} \! \text{Im}\( G(\v e_{12; 34}) G(\v e_{56; 34}) \lam_{1234}\lam_{5612}\lam_{3456}\) (\delta_{1r}{+}\delta_{2r} {-}\delta_{3r} {-} \delta_{4r})\\
\prod_{i=1}^6 n_i
\Big(\frac{1}{n_1}{+}\frac{1}{n_2} \Big) 
\Big(\frac{1}{n_3}{+}\frac{1}{n_4} {-}\frac{1}{n_5} {-}\frac{1}{n_6}\Big)~.
\end{multline}
The other term, $\int d J\, J_r\,  \(\delta \mathcal{L}\)_{\text{second}}^b$, is obtained by the transformation discussed previously:  $2\rightarrow - 3$, $3 \rightarrow - 2$, and $6\rightarrow - 6$, and multiplying by a combinatorial factor of $4$, giving, 
\bml
\int d J\, J_r\,  \(\delta \mathcal{L}\)_{\text{second}}^b= 16 i\sum_{1,\ldots, 6} \text{Im}\( G(\v e_{12; 34}) G(\v e_{52;46}) \lam_{1234}\lam_{5316}\lam_{4625}\) (\delta_{1r}{+}\delta_{2r} {-}\delta_{3r} {-} \delta_{4r})\\
\prod_{i=1}^6 n_i
\Big(\frac{1}{n_1}{-}\frac{1}{n_3} \Big) 
\Big(\frac{1}{n_4}{+}\frac{1}{n_6} {-}\frac{1}{n_2} {-}\frac{1}{n_5}\Big)~.
\end{multline}

Combining all these terms gives, via (\ref{415}), the next-to-leading order kinetic equation. We would like to make sure that it matches the result we found previously in \cite{RS1}. Focusing on (\ref{49}), we see that we can replace $(\delta_{1r}{+}\delta_{2r} {-}\delta_{3r} {-} \delta_{4r})\rightarrow 2(\delta_{1r}{-}\delta_{3r} )$ because of the symmetry of everything else under $1\leftrightarrow 2$ or $3 \leftrightarrow 4$. For the term coming with $\delta_{3r}$ we do a change of variables, $(1,2) \leftrightarrow (3,4)$, 
\bml \label{4200}
\int d J\, J_r\,  \(\delta \mathcal{L}\)_{\text{second}}^a\\= -8i\sum_{1,\ldots, 6} \delta_{1r}\prod_{i=1}^6 n_i  \Big\{\text{Im} \( G(\v e_{12; 34}) G(\v e_{56; 34}) \lam_{1234}\lam_{5612}\lam_{3456} \)\Big(\frac{1}{n_1}{+}\frac{1}{n_2}\Big)
\Big(\frac{1}{n_3}{+}\frac{1}{n_4}{-}\frac{1}{n_5}{-}\frac{1}{n_6}\Big)\\
+\text{Im} \( G(\v e_{12; 34}) G(\v e_{ 12;56}) \lam_{1234}\lam_{5612}\lam_{3456} \)\Big(\frac{1}{n_3}{+}\frac{1}{n_4}\Big)
\Big(\frac{1}{n_1}{+}\frac{1}{n_2}{-}\frac{1}{n_5}{-}\frac{1}{n_6}\Big)\Big\}~,
\end{multline}
where we used that $G(- \v e) =-G^*(\v e)$. 
Next, we insert the following relation,
\be
G(\v e_{12; 34}) (G(\v e_{56; 34})+ G(\v e_{ 12;56})) =G(\v e_{56; 34})G(\v e_{ 12;56}) 
\ee
into (\ref{4200}).  Writing the $G(\v e_{ij;kl})$ in (\ref{4200}) explicitly, we have, 
\bml 
\!\!\!\!\!\int d J\, J_r\,  \(\delta \mathcal{L}\)_{\text{second}}^a= -8i\sum_{1,\ldots, 6} \delta_{1r} \prod_{i=1}^6 n_i 
 \text{Im}\Big[\lam_{1234}\lam_{5612}\lam_{3456} \Big\{\Big(\frac{1}{n_1} {+} \frac{1}{n_2} \Big)\Big(\frac{1}{n_3} {+} \frac{1}{n_4} \Big)\frac{1}{\o_{34;56}{-}i\eps}\frac{1}{\o_{56;12} {-}i \eps} \\- \Big(\frac{1}{n_5} {+} \frac{1}{n_6} \Big)\Big(\frac{1}{n_1} {+} \frac{1}{n_2} \Big) \frac{1}{\o_{34;12}{-}i\eps}\frac{1}{\o_{34;56} {-}i \eps} - \Big(\frac{1}{n_5} {+} \frac{1}{n_6} \Big)\Big(\frac{1}{n_3} {+} \frac{1}{n_4} \Big) \frac{1}{\o_{34;12}{-}i\eps}\frac{1}{\o_{56;12} {-}i \eps}  \Big\}\Big]~,
\end{multline}
where we defined the shorthand $\o_{ij; kl} \equiv \o_i {+} \o_j {-} \o_k {-} \o_l$. This precisely reproduces what we had in the kinetic equation in \cite{RS1}.~\footnote{ In particular,  in \cite{RS1} insert (4.25)  into (4.32) and use that the $\text{Im}(x^*) = - \text{Im}(x)$. } There we proceeded by using (\ref{Gsplit}) on all the $G(\v e_{ij;kl})$ terms, grouping terms together based on the number of delta functions of the $\o_i$, and performing the change of variables $(3,4) \leftrightarrow (5,6)$ on some of the terms in order to show that they are either zero due to antisymmetry, or to combine  them with other terms. The end result is the kinetic equation (\ref{KEQ}).

\section{Frequency renormalization}
In the Hamiltonian (\ref{33}), 
 \be 
 H = \sum_i \o_i J_i + \sum_{i,j,k,l} \lam_{ijkl} \sqrt{J_i J_j J_k J_l} \exp\( i (\al_i{ +} \al_j {-} \al_k{ -} \al_l)\)~,
 \ee
 it is useful to split off the term in the interaction in which some pair of indices are equal, 
  \bea
 H &=& \sum_i \o_i J_i + \sum_{(i,j) \neq (k,l)} \lam_{ijkl} \sqrt{J_i J_j J_k J_l} \exp\( i (\al_i{ +} \al_j {-} \al_k{ -} \al_l)\) + H_{ren}~,  \
 \\ 
 H_{ren}&=&  2\sum_{i \neq k} \lam_{i k i k} J_i J_k  
  + \sum_i \lambda_{iiii} J_i^2  ~.
\nonumber 
\eea
 The corresponding matrix element of $(\delta L)_{ren}$ for $H_{ren}$ is, from (\ref{Lio2}), 
 \be
 \la \v n| (\delta L)_{ren}|\v n'\ra =  4 \,\delta_{\v n, \v n'} \sum_{i \neq k } \lam_{i k ik} J_k n_i  +   2 \,\delta_{\v n, \v n'}  \sum_i \lambda_{iiii} J_i n_i  ~.
 \ee
Switching to notation in which we call mode $i$ to be $1$ and mode $k$ to be $5$, we have that the contribution at order $\lam^3$ which involves one insertion of $(\delta L)_{ren}$ is, 
\bml
\(\delta \mathcal{L}\)_{\text{ren}} =-  4 \sum_{1, \ldots, 5}  G(\v e_{12;34})^2 |\lam_{1234}|^2  \(\d_1{+}\d_2{-}\d_3{-}\d_4\) J_1 J_2  J_3 J_4
\\
 J_5 \( \lam_{1515} {+}\lam_{2525} {-}\lam_{3535} {-}\lam_{4545}\)
\(\d_3{+}\d_4{-}\d_1{-}\d_2\) \la 0 |\Phi_0\ra~,
\end{multline}
where we have accounted for the fact that $(\delta L)_{ren}$ can be inserted along any of the lines $1,2,3,4$.
The contribution to the kinetic equation is, 
\bml \label{B5}
\int d J\, J_r\,  \(\delta \mathcal{L}\)_{\text{ren}}  = i\sum_{1,\ldots, 4}\text{Im}(G(\v e_{12; 34})^2) |\lam_{1234}|^2(\delta_{1r}{+}\delta_{2r} {-}\delta_{3r} {-} \delta_{4r}) \prod_{i=1}^4 n_i\,
\Big(\frac{1}{n_1}{+}\frac{1}{n_2} {-}\frac{1}{n_3} {-}\frac{1}{n_4}\Big)\\
 \( \delta \o_1 {+}\delta \o_2 {-} \delta \o_3 {-} \delta \o_4\)
\end{multline}
where we used that the term with $\text{Re}(G(\v e_{12; 34})^2)$ gives zero due to the symmetry under $1,2\leftrightarrow 3,4$, and where
\be
\delta \o_a =   4  \sum_5 \lam_{a5 a 5}  n_5~.
\ee
We note that, 
\be
G(\v e_{12;34}) = \frac{1}{\o_{12;34} {+} i \eps}~, \ \ \ \Rightarrow \, \, \text{Im}(G(\v e_{12; 34})^2)  = -\frac{\d}{\d \o_{12;34}}\text{Im}  \frac{1}{\o_{12;34} {+} i \eps} = \pi \delta'(\o_{12;34})
\ee
The contribution to the kinetic equation of (\ref{B5}) is therefore, 
\be \label{B8}
\(\frac{d n_r}{\d t}\)_{\text{ren}} \!\!\!=  4\sum_{1,\ldots, 4}\pi \delta'(\o_{12;34}) |\lam_{1234}|^2(\delta_{1r}{+}\delta_{2r} {-}\delta_{3r} {-} \delta_{4r}) \prod_{i=1}^4 n_i\,
\Big(\frac{1}{n_1}{+}\frac{1}{n_2} {-}\frac{1}{n_3} {-}\frac{1}{n_4}\Big)\\
 \( \delta \o_1 {+}\delta \o_2 {-} \delta \o_3 {-} \delta \o_4\)
 \ee
 The reason we interpreted this as a frequency renormalization is because if one takes the leading order kinetic equation, 
\be
\frac{\d n_r}{\d t} = 4 \sum_{1, \ldots, 4}(\delta_{1r}{+}\delta_{2r} {-}\delta_{3r} {-} \delta_{4r})  \pi \delta(\o_{12;34})| \lam_{1234}|^2  \prod_{i=1}^4 n_i\, \Big( \frac{1}{n_1} {+} \frac{1}{n_2}{-}\frac{1}{n_3} {-} \frac{1}{n_4} \Big) 
\ee
and sets $\o_a \rightarrow \o_a + \delta \o_a$, the change is precisely (\ref{B8}).

\end{document}